\documentclass[]{aipproc}

\layoutstyle{8x11single}

\begin{document}

\title{Hydrodynamical analysis of flow at RHIC}

\classification{25.75-q, 25.75.Ld}
\keywords{Relativistic heavy-ion collisions; Elliptic flow; 
          Hydrodynamic model}

\author{Pasi Huovinen}{
  address={Lawrence Berkeley National Laboratory, Berkeley, USA},
  email={phuovinen@lbl.gov}
}

\iftrue
%
\fi

\copyrightyear  {2001}

\begin{abstract}
We use a hydrodynamical model to describe the evolution of the collision system
at collision energies $\sqrt{s}=130$ and 200 GeV. At lower $\sqrt{s}=130$ GeV
energy we compare the results obtained assuming fast or slow thermalization
(thermalization times $\tau_0=0.6$ and 4.1 fm/$c$, respectively) and show that
slow thermalization fails to reproduce the observed anisotropy of particle
distribution. At $\sqrt{s}=200$ GeV collision energy our results show
anisotropies similar to those observed at $\sqrt{s}=130$ GeV.

\end{abstract}

\date{\today}

\maketitle

\section{Introduction}

The azimuthal anisotropy of particle distributions in heavy ion collisions
is a signal of rescattering among the particles. The larger the anisotropy,
the stronger the rescattering. Since the hydrodynamical models assume zero
mean free path and thus infinite scattering rate, they provide an upper limit
to observable anisotropies. In our earlier
papers~\cite{Kolb:2001,Huovinen:2001,Kolb:2001b} we have argued that since the
observed anisotropies are well reproduced by a hydrodynamical model, the
collision system must thermalize fast. Here we constrain our model further
by requiring it to reproduce the observed single particle
$p_t$-distributions~\cite{Velkovska:2001} and study the consequences of such a
long thermalization time that there is no thermal plasma but the hydrodynamical
expansion begins in hadronic phase. We also tune our model to reproduce 
the charged particle multiplicity at $\sqrt{s}=200$ GeV collision energy
published recently by the PHOBOS collaboration~\cite{Phobos} and show our
results for the charged particle anisotropy and pion and antiproton
$p_t$-spectra at the higher energy.

\section{Hydrodynamical model and its initialization}

To reduce the complexity of the numerical calculations we assume boost
invariant scaling flow in our hydrodynamical model and solve the equations 
numerically only in the two transverse dimensions. We also apply ideal fluid
hydrodynamics and assume that viscous effects are negligible. Depending on the
initialization we apply two different equations of state: one with 
a phase transition at $T_c=165$ MeV or purely hadronic equation of
state, called EoS\,A and H in~\cite{Kolb:2001}, respectively.

In our model we assume that the chemical and kinetic freeze-out take place at
the same temperature. In other words the system maintains both kinetic and
chemical local equilibrium during the entire evolution from thermalization to
freeze-out. This makes it impossible for us to fit both proton and antiproton
yields simultaneously since the thermal model studies~\cite{Xu:2001} point to
chemical freeze-out temperatures much higher than the kinetic freeze-out
temperature $T_f = 100$ -- 140 MeV required to produce sufficiently flat
$p_t$-spectra. We counter this problem by the following approximation: We fix
our initial baryon density to reproduce the observed
$\bar{p}\slash p \sim 0.6$~\cite{Back:2001} 
ratio at $T=165$ MeV, i.e.\ immediately after the system has
hadronized. We then let the system evolve to its freeze-out temperature
and scale the calculated proton and antiproton yields to their values at
$T=165$ MeV by hand. This scaling factor depends on the kinetic freeze-out
temperature and has to be adjusted accordingly.

	\subsection{Fast thermalization}

We have discussed different ways to parametrize the initial state of
hydrodynamic evolution in~\cite{Kolb:2001b}. It was found that neither a
parametrization where the density (energy or entropy) is proportional to the
local number of participants or binary collisions is sufficient to reproduce
the observed centrality dependence of multiplicity~\cite{Adcox:2001}.
A combination of these like the two component model presented
in~\cite{Kharzeev:2001} is required instead.

To obtain the results presented here we have used a combination of
parametrizations eWN and eBC described in~\cite{Kolb:2001b}\footnote{
      A combination of sWN and sBC of~\cite{Kolb:2001b} is equally possible.}.
In this case we assume the initial energy density to be proportional to the
scaled sum of number of participants and binary collisions per unit area: 
\begin{equation}
  e(\mathbf{s},\tau_0;\mathbf{b})
   = x K_e(\tau_0)n_\mathrm{WN}(\mathbf{s;b})
    +(1-x) \tilde{K}_e(\tau_0)n_\mathrm{BC}(\mathbf{s;b}),
  \label{initial1}
\end{equation}
where $n_\mathrm{WN}(\mathbf{s;b})$ is the number of participants and
$n_\mathrm{BC}(\mathbf{s;b})$ the 
number of binary collisions per unit area in the transverse plane. The
normalization constants $K_e(\tau_0)$ and $\tilde{K}_e(\tau_0)$ are the same
than in~\cite{Kolb:2001b} and the mixing parameter $x=0.4$ was found to
reproduce the observed centrality dependence of multiplicity. The
equation of state with phase transition (EoS\,A) has been used to calculate
results with short thermalization time ($\tau_0=0.6$ fm).

	\subsection{Slow thermalization and free streaming}

The parametrization described above is
reasonable if thermalization time is sufficiently small and produced particles
do not have time to move far away from the place where they were produced.
On the other hand, if it takes several fermi for the system to thermalize, the
particles will move during thermalization and smear any structure in the
system.

To estimate what the shape and distributions of the system might be after a 
long thermalization process, we developed further the ideas
outlined in~\cite{Kolb:2000}. We approximate the expansion of the system
during thermalization by free streaming. In this approach the number density
of initially produced excitations is again taken to be proportional to the
scaled sum of number of participants and binary collisions whereas the
momentum distribution is taken to be that of pions at temperature of
$T=200$ MeV. The free streaming is allowed to continue until thermalization
time $\tau_0$ when energy density distribution due to free streaming particles
is calculated and taken to be the thermal energy distribution. In this
approximation we do not assign any radial velocity to the system at the
end of free streaming. The effects of build-up of flow during thermalization
will be discussed in~\cite{KTHH}.

\begin{figure}[b]
 \begin{minipage}[t]{75mm}
  \resizebox{54mm}{!}{\includegraphics{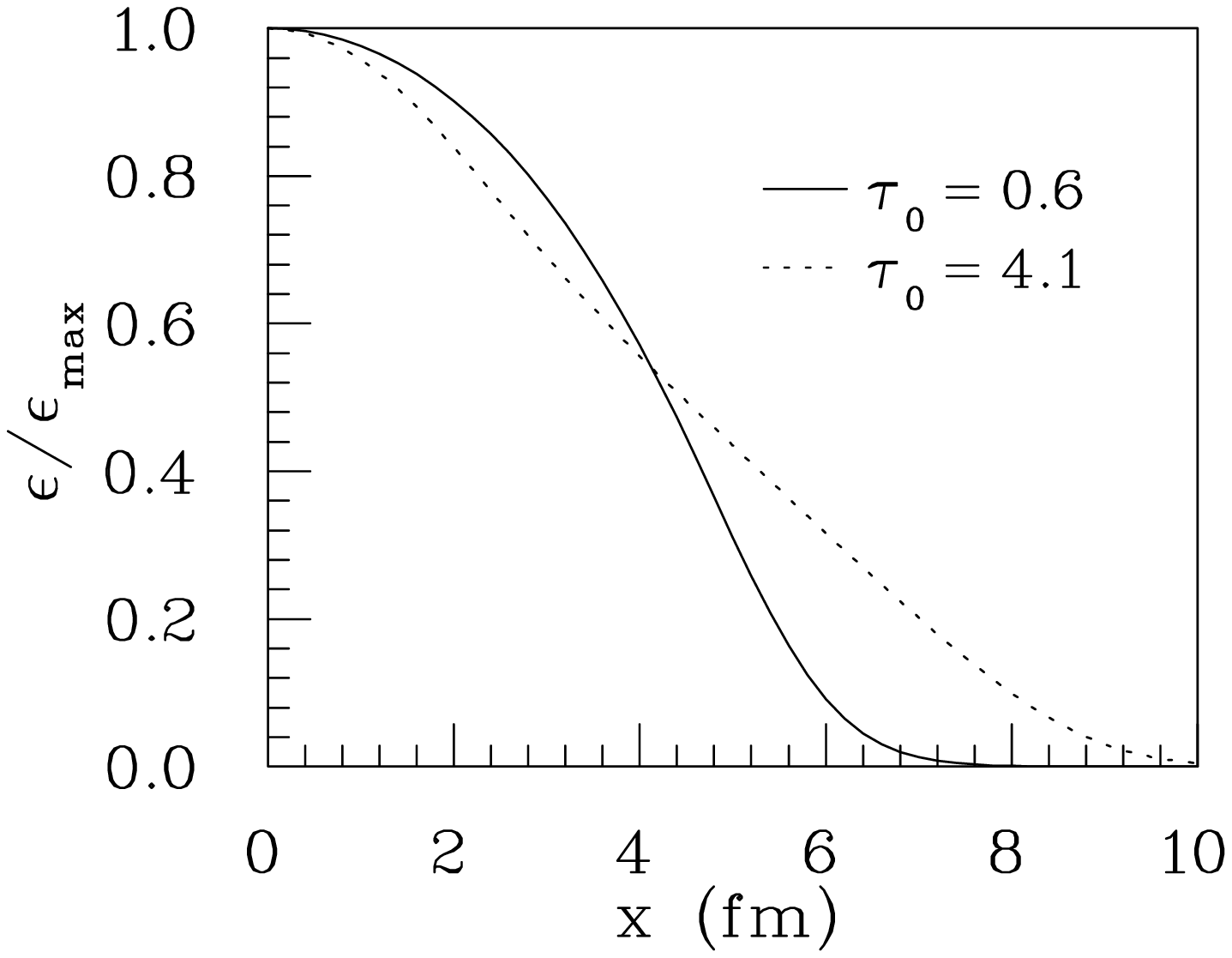}}
 \end{minipage}
    \hfill
 \begin{minipage}[t]{75mm}
  \resizebox{52mm}{!}{\includegraphics{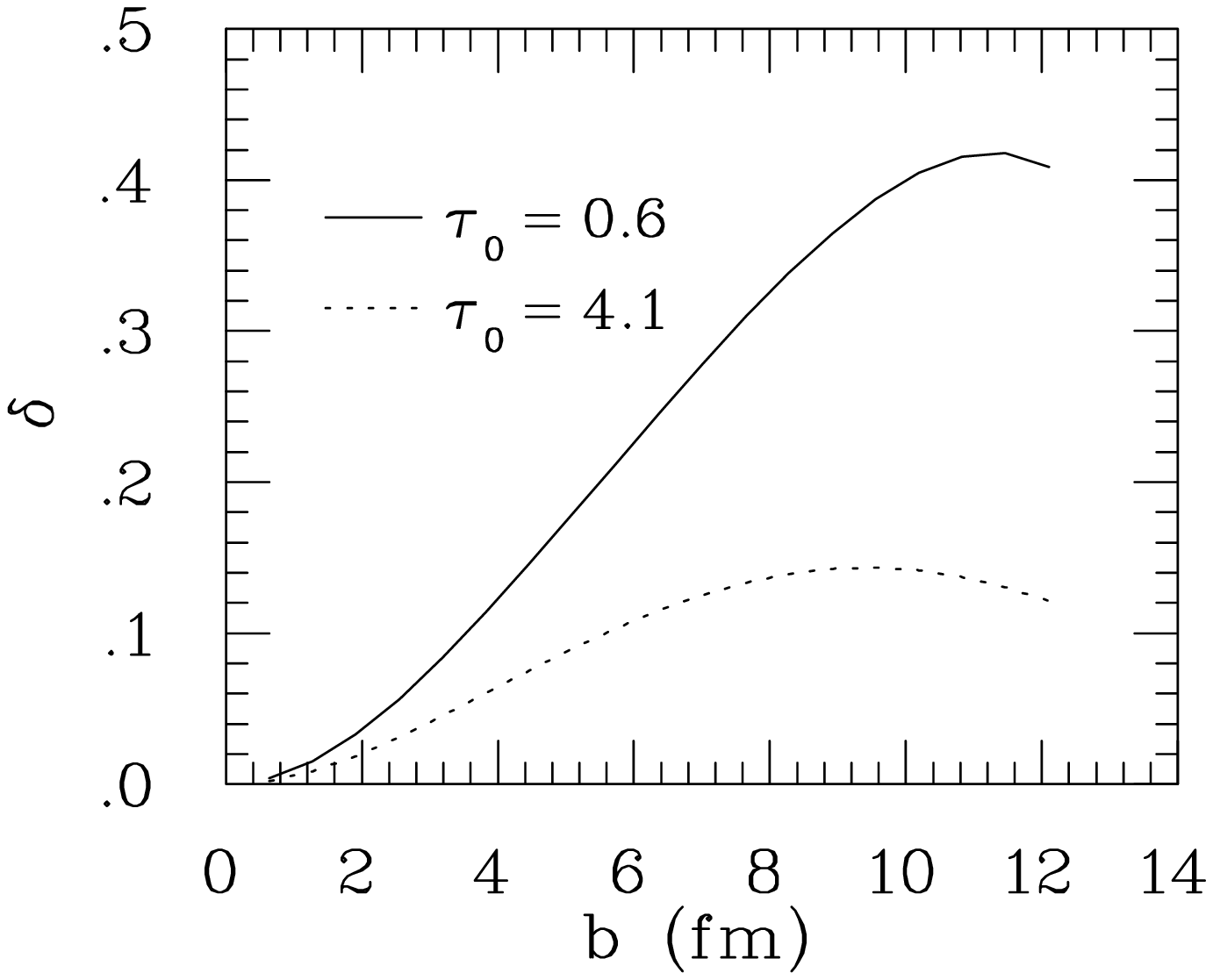}}
    \caption{Initial energy density distribution in $b=2.4$ fm collision
             scaled by maximum energy density (left) and initial spatial
             deformation of the system as function of impart parameter (right)
             using parametrizations for fast and slow thermalization.}
 \end{minipage}
 \label{shapes}
\end{figure}

The distribution of particles in a free streaming system is given
by~\cite{Heiselberg:1999}
\begin{equation}
  n(\mathbf{x,p},\tau_0) = 
    f_0(p_t,p'_z)\,S_\perp(\mathbf{s-v}_\perp(\tau_0-\tau_i),\mathbf{b}),
  \label{initial2}
\end{equation}
where $S_\perp(\mathbf{s,b}) \propto x K_n n_\mathrm{WN}(\mathbf{s;b})
                      + (1-x) K_b n_\mathrm{BC}(\mathbf{s;b})$ gives the
density distribution of the initial excitations and $f_0(p_t,p_z)$ is Boltzmann
distribution. Of the parameters $p'_z = \tau_0 p_z\slash\tau_i$ is the scaled
longitudinal momentum which takes longitudinal free streaming in boost
invariant system into account, $\tau_0$ is thermalization time and $\tau_i$ is
the time when the distributions before free streaming are specified. We use
$\tau_i = 0.6$ fm and fix thermalization time by requiring that maximum
temperature in central collisions is low enough for hadrons to exist. We choose
this somewhat arbitrary limit to be $T=200$ MeV. This corresponds to
$\epsilon = 2.3$ GeV/fm$^3$ in the hadronic equation of state (EoS H) we use
with this initial state. After the remaining parameters $K_n$, $K_b$ and $x$
are chosen to reproduce the observed centrality dependence of
multiplicity~\cite{Adcox:2001}, this procedure leads to a thermalization time
$\tau_0=4.1$ fm. 

As shown in Fig.~\ref{shapes} this parametrization leads to a larger system
which density gradients are flatter and the spatial deformation
smaller than in the fast thermalization parametrization (Eq.(\ref{initial1})).

\section{Comparison to observables}

	\subsection{$p_t$-distributions}

\begin{figure}
 \begin{minipage}[t]{75mm}
  \resizebox{65mm}{!}{\includegraphics{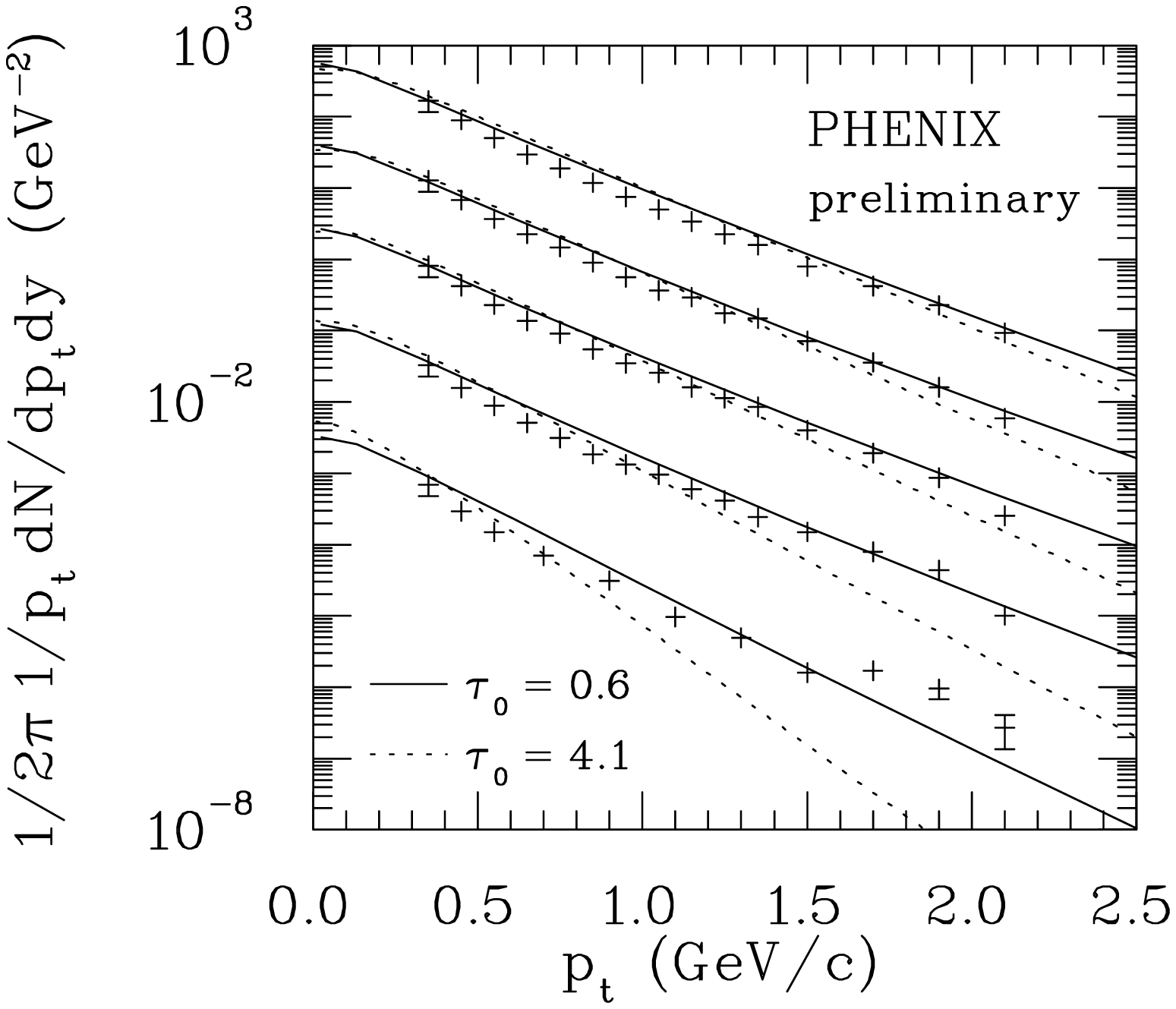}}
 \end{minipage}
    \hfill
 \begin{minipage}[t]{75mm}
  \resizebox{65mm}{!}{\includegraphics{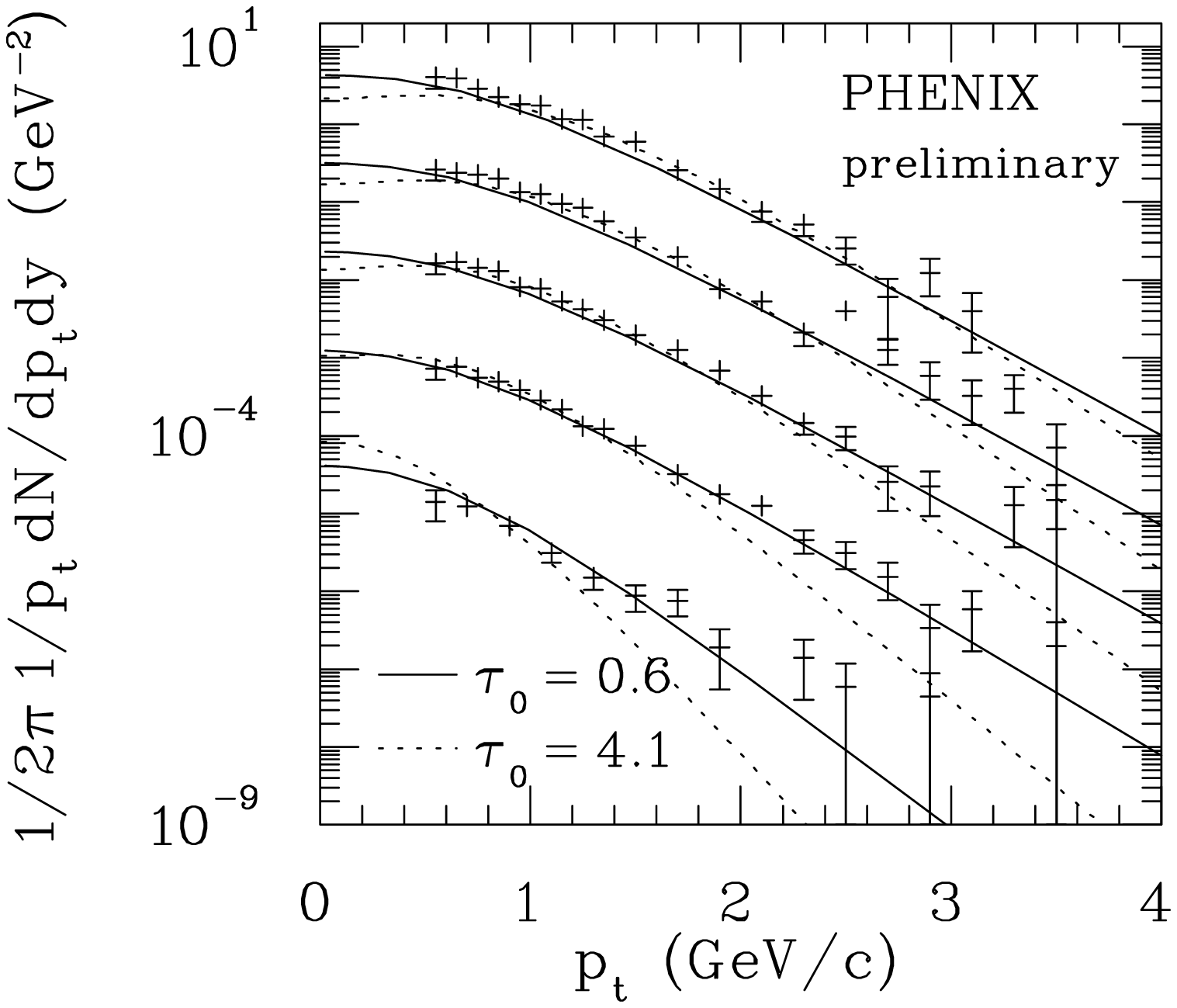}}
    \caption{$p_t$ distributions of negative pions (left) and antiprotons
             (right) at $\sqrt{s}=130$ collision energy compared with
             preliminary PHENIX data~\cite{Velkovska:2001}. The data and curves
             correspond to centralities
             $\sigma\slash\sigma_\mathrm{tot}=0$-5\%, 5-15\%, 15-30\%, 30-60\%,
             60-92\% and impact parameters $b=2.4$, 5.1, 7.0, 9.6 and 12.1 fm,
             respectively (top to bottom). These data sets are successively
             scaled down by $10^{-n}$,$n=0$,1,2,3,4.}
 \end{minipage}
 \label{pts130}
\end{figure}

After fixing the proportionality constants $K_e$ and $\tilde{K}_e$ and mixing
$x$ in Eq.(\ref{initial1}) the only remaining parameter in our model is the
freeze-out temperature. We fix it to reproduce the preliminary
$p_t$-distributions of pions and antiprotons in most central collisions 
measured by the PHENIX Collaboration~\cite{Velkovska:2001}. This
requirement leads to very different temperatures. If early thermalization is
assumed, $T_f\approx 140$ MeV leads to a nice fit. On the other hand long
thermalization time and free streaming of particles during thermalization
smoothens the pressure gradients so much that a low freeze-out temperature of
$T_f\approx 100$ MeV is needed to create large enough transverse flow
(Fig.(\ref{pts130})).
This difference in freeze-out temperatures leads also to very different
scaling factors for the antiproton yield, 2 and 27 for $\tau_0 = 0.6$ fm and
$\tau_0=4.1$ fm, respectively. After scaling the yields to their values at
$T_{ch}=165$ MeV, we reach an acceptable reproduction of measured spectra in
most central collisions.

Having the parameters of our models fixed we can now compare our results to
spectra of non-central collisions~\cite{Velkovska:2001} (see
Fig.~\ref{pts130}). The average number of
participants in five different centrality classes~\cite{Burward-Hoy} set our
choice of impact parameter in these bins to 2.4, 5.1, 7.0, 9.6 and 12.1 fm,
respectively. The early thermalization assumption leads to very good agreement
with the preliminary data even for the most peripheral collisions. On the other
hand, the late thermalization approach performs adequately in central and
semi-central collisions, but leads to far too steep slopes in peripheral
collisions. This failure could be expected since if one assumes a long
thermalization time even in central collisions, it is probable that the system
does not thermalize at all in peripheral collisions and our model fails.

	\subsection{Azimuthal anisotropy}

\begin{figure}
  \resizebox{65mm}{!}{\includegraphics{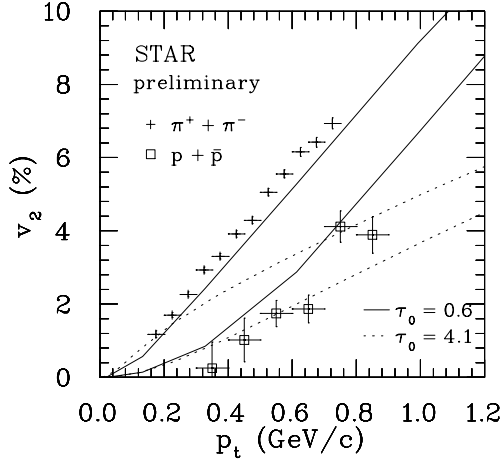}}
 \caption{Differential anisotropy of pions and protons+antiprotons (upper and
          lower curve, respectively) in minimum bias collisions compared to
          preliminary data~\cite{Adler:2001}.}
 \label{v2}
\end{figure}

So far we have used different observables to fix parameters of our model and
the real test is in the comparison to the observed anisotropies of particle
distribution~\cite{Adler:2001,Snellings:2001}. As we have reported in our
earlier papers~\cite{Kolb:2001,Huovinen:2001,Kolb:2001b}, hydrodynamical
model leads to an excellent
fit to observed data if thermalization time is short. We show our present
results in Fig.(\ref{v2}). The present parametrization ($\tau_0=0.6$ fm)
leads to
a slightly smaller differential anisotropy for pions and larger anisotropy for
antiprotons than shown in our previous works. Both changes are mainly due to
higher freeze-out temperature (see~\cite{Huovinen:2001}).
The deviation from data is still within systematical error and as reported in
this conference~\cite{Tang}, more careful analysis of the data to make
systematic errors smaller may lead to slightly smaller $v_2$. Also we want to
remind that our present fit to freeze-out temperature $T_f\approx 140$ MeV is
preliminary and may change when we fit more data and explore different
parameter combinations.

On the other hand the results for long thermalization time approach are well
below the data and show that -- at least in the present formulation -- the free
streaming period changes the shape of the initial system sufficiently to
prevent the build-up of observed anisotropies. This corroborates
our earlier argument that early thermalization is necessary to achieve the
observed large anisotropies in particle distributions.

\section{Collisions at $\sqrt{s}=200$ GeV}

\begin{figure}[b]
 \begin{minipage}[t]{75mm}
  \resizebox{67mm}{!}{\includegraphics{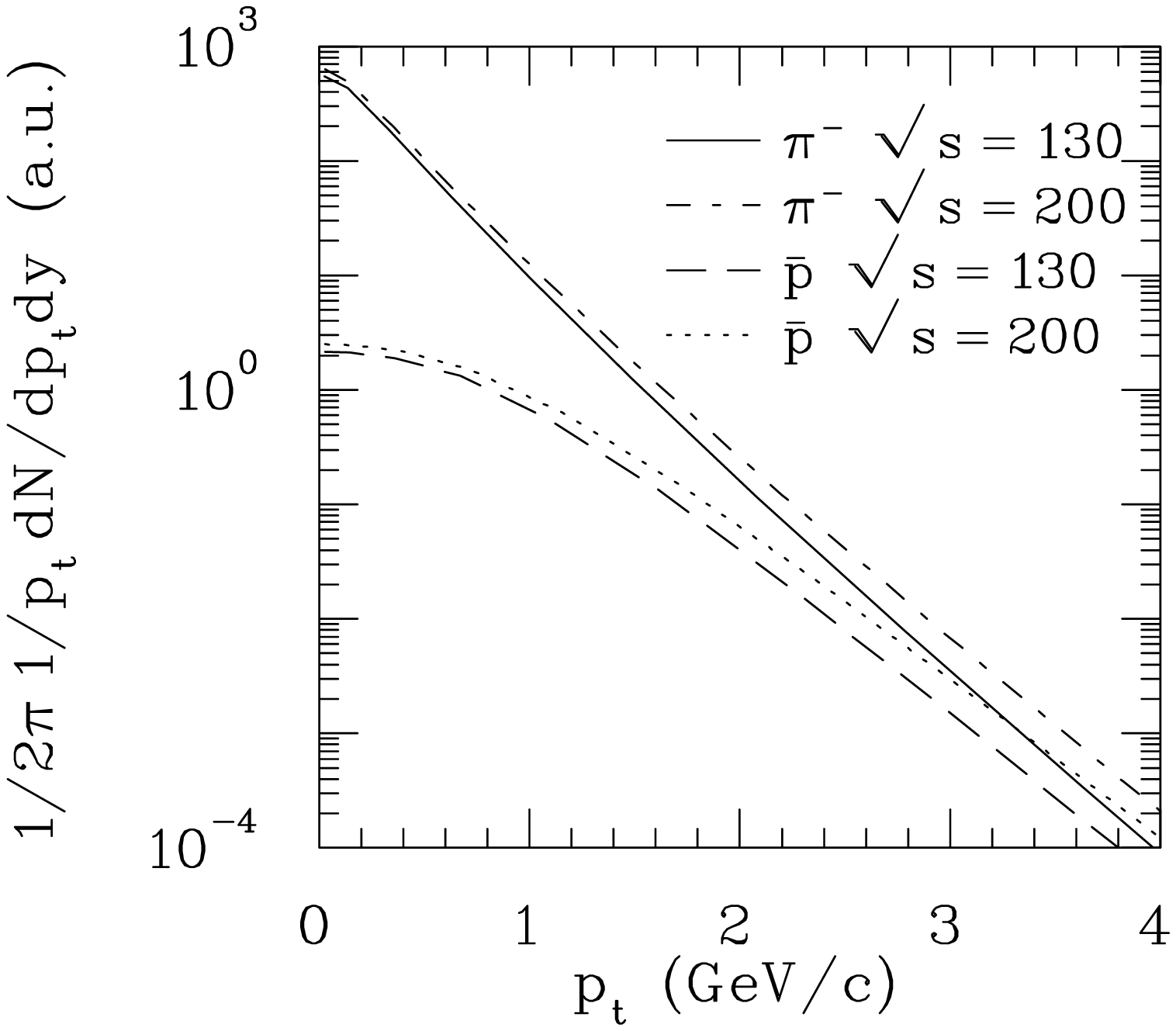}}
 \end{minipage}
    \hfill
 \begin{minipage}[t]{75mm}
  \resizebox{60mm}{!}{\includegraphics{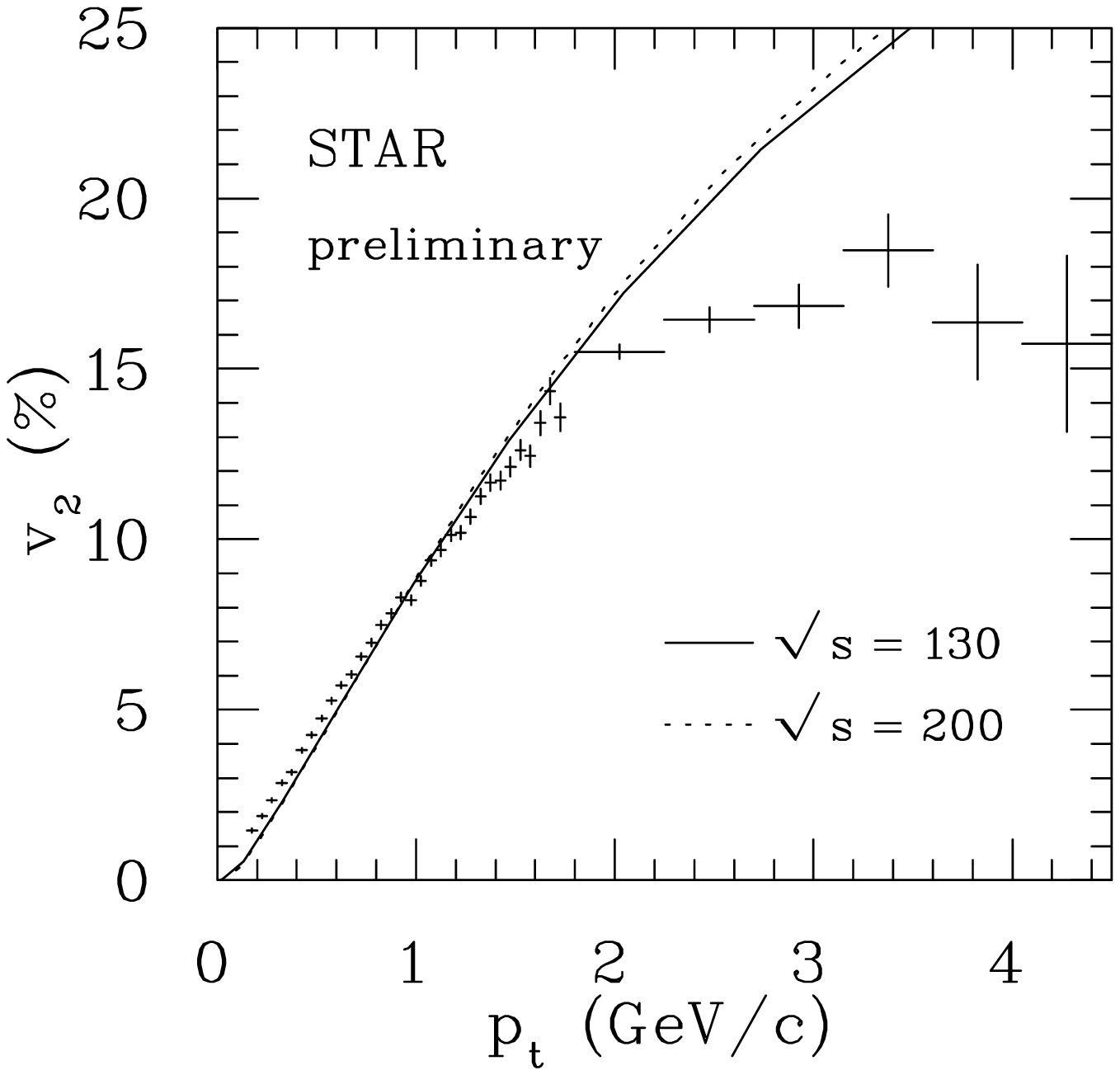}}
    \caption{$p_t$ distributions of pions and antiprotons at 5\% most central
             $\sqrt{s}=130$ and 200 GeV collisions (left)
            and differential anisotropy of charged particles in minimum bias
            $\sqrt{s}=130$ and 200 GeV collisions compared with preliminary
            STAR data at $\sqrt{s}=130$ GeV~\cite{Snellings:2001} (right).}
 \end{minipage}
 \label{200GeV}
\end{figure}

We use the charged particle multiplicity $dN\slash d\eta = 650 \pm 35$ in
6\% most central collisions measured recently by the PHOBOS
Collaboration~\cite{Phobos} to fix our parametrization for $\sqrt{s}=200$ GeV
collisions. We assume that all the other parameters like freeze-out temperature
($T_f\approx 140$ MeV) and thermalization time ($\tau_0=0.6$ fm) stay
unchanged. We do not have any method to calculate the change in baryon stopping
and thus the change in initial baryon density. Therefore we ignore the question
of the change in stopping altogether and use the same initial baryon density at
both energies. For the same reason we do not scale the calculated antiproton
yield at freeze-out either but show the result as it is to demonstrate the
change in slopes.

The calculated pion and antiproton distributions at $\sqrt{s}=130$ and 200
GeV are shown in Fig.(\ref{200GeV}). The changes in slopes are small but still
large enough to increase $E_t$ by $\sim 18\%$ even if the observed increase in
particle multiplicity is $14 \%$~\cite{Phobos}. Our calculation leads to the
increase of the average energy density at $\tau=1$ fm from 5.4 to 6.8
GeV/fm$^3$ and the increase of the maximum temperature (at $\tau_0 = 0.6$ fm)
from 355 to 375 MeV. If the changes in $p_t$-spectra are small they are even
tinier in anisotropies. In Fig.(\ref{200GeV}) differential anisotropy for
charged particles in minimum bias collisions is shown with preliminary STAR
data for $\sqrt{s}=130$ GeV collision. 
The change is negligible and it will be interesting to see
whether the data deviates from the calculation around $p_t \approx 2$ GeV
also at $\sqrt{s}=200$ GeV collision energy.

\section{Summary}

We have shown that it is possible to fit both the experimental $p_t$ spectra
and anisotropy of particle distribution if short thermalization time and
hydrodynamical behaviour of the collision system is assumed. On the other hand
long thermalization time allows the system shape change too much before the
build-up of flow begins and generates sufficient anisotropies. This
corroborates our earlier argument that fast thermalization is required to
explain the experimental data.

We have also made the first calculations at $\sqrt{s}=200$ GeV collision
energy. Our results show only slightly flatter spectra and basically no change
in differential $v_2$ when compared to results at lower $\sqrt{s}=130$ GeV
energy.

\begin{theacknowledgments}
This work has been done in collaboration with P.F.~Kolb and U.~Heinz.
The discussions with and ideas given by Henning Heiselberg and Peter Jacobs
are gratefully acknowledged. This work was partly supported by the
U.S. Department of Energy under Contract No.\ DE-AC03-76SF00098.
\end{theacknowledgments}



\begin{thebibliography}{19}

\bibitem{Kolb:2001}
P.~F.~Kolb, P.~Huovinen, U.~Heinz and H.~Heiselberg,
Phys.\ Lett.\ B {\bf 500}, 232 (2001)
[hep-ph/0012137].

\bibitem{Huovinen:2001}
P.~Huovinen, P.~F.~Kolb, U.~Heinz, P.~V.~Ruuskanen and S.~A.~Voloshin,
Phys.\ Lett.\ B {\bf 503}, 58 (2001)
[hep-ph/0101136].

\bibitem{Kolb:2001b}
P.~F.~Kolb, U.~Heinz, P.~Huovinen, K.~J.~Eskola and K.~Tuominen,
hep-ph/0103234.

\bibitem{Velkovska:2001}
J.~Velkovska  [PHENIX collaboration], nucl-ex/0105012.

\bibitem{Phobos}
B.~Wyslouch, these proceedings;
B.~B.~Back {\it et al.}  [PHOBOS Collaboration], nucl-ex/0108009.

\bibitem{Xu:2001}
N.~Xu and M.~Kaneta, nucl-ex/0104021;
P.~Braun-Munzinger, D.~Magestro, K.~Redlich and J.~Stachel, hep-ph/0105229;
W.~Florkowski, W.~Broniowski and M.~Michalec, nucl-th/0106009.

\bibitem{Back:2001}
B.~B.~Back {\it et al.}  [PHOBOS Collaboration], hep-ex/0104032;
C.~Adler  [the STAR Collaboration],
Phys.\ Rev.\ Lett.\  {\bf 86}, 4778 (2001) [nucl-ex/0104022];
I.~G.~Bearden {\it et al.}  [BRAHMS Collaboration], nucl-ex/0106011.

\bibitem{Adcox:2001}
K.~Adcox {\it et al.}  [PHENIX Collaboration],
Phys.\ Rev.\ Lett.\  {\bf 86}, 3500 (2001) [nucl-ex/0012008];
B.~B.~Back {\it et al.}  [PHOBOS Collaboration], nucl-ex/0105011.

\bibitem{Kharzeev:2001}
D.~Kharzeev and M.~Nardi, Phys.\ Lett.\ B {\bf 507}, 121 (2001)
[nucl-th/0012025].

\bibitem{Kolb:2000}
P.~F.~Kolb, J.~Sollfrank and U.~Heinz, 
Phys.\ Rev.\ C {\bf 62}, 054909 (2000) [hep-ph/0006129].

\bibitem{KTHH}
P.~F.~Kolb, M.~Tilley, P.~Huovinen, and U.~Heinz in preparation.

\bibitem{Heiselberg:1999}
H.~Heiselberg and A.~Levy,
Phys.\ Rev.\ C {\bf 59}, 2716 (1999)
[nucl-th/9812034];
G.~Baym, Phys.\ Lett.\ B {\bf 138}, 18 (1984).

\bibitem{Burward-Hoy}
J.~Burward-Hoy, private communication.

\bibitem{Adler:2001}
C.~Adler {\it et al.}  [STAR Collaboration], nucl-ex/0107003.

\bibitem{Snellings:2001}
R.~J.~Snellings  [STAR Collaboration], nucl-ex/0104006.

\bibitem{Tang}
A.H.~Tang, these proceedings.

\end{thebibliography}
\end{document}